\documentclass[twocolumn,prl,notitlepage,floats,superscriptaddress,amsmath,amssymb]{revtex4-2}

\usepackage[version=3]{mhchem} 
\usepackage{bm}
\usepackage[utf8]{inputenc}
\usepackage[T1]{fontenc}
\usepackage{graphicx}
\usepackage{upgreek}
\usepackage{color}
\usepackage{ulem}
\usepackage{hyperref} 
\hypersetup{colorlinks,citecolor=blue, filecolor=blue ,linkcolor=blue , urlcolor=blue, pdftex}
\usepackage{sidecap}
\usepackage{amssymb}
\usepackage[caption=false]{subfig}

\def \FUW{Institute of Experimental Physics, Faculty of Physics, University of Warsaw, 02-093 Warsaw, Poland}

\def \Watanabe{Research Center for Functional Materials, National Institute for Materials Science, Tsukuba 305-0044, Japan}
\def \Taniguchi{International Center for Materials Nanoarchitectonics, National Institute for Materials Science, Tsukuba 305-0044, Japan}
\def \Sapienza{Physics Department, Sapienza University of Rome, 00185 Rome, Italy}
\def \CNR{Institute for Photonics and Nanotechnologies, National Research Council (CNR-IFN), 00133 Rome, Italy}
\def \LNCMI{Laboratoire National des Champs Magnétiques Intenses, CNRS-UGA-UPS-INSA-EMFL, 38042 Grenoble, France}
\def \Brno{Central European Institute of Technology, Brno University of Technology, Brno, Czech 61200, Republic}
\def \Centera{CENTERA Laboratories, Institute of High Pressure Physics, Polish Academy of Sciences, 01-142 Warsaw, Poland}

\begin{document}

\title{Excitons and trions in WSSe monolayers}

\author{Katarzyna Olkowska-Pucko}
\email{katarzyna.olkowska-pucko@fuw.edu.pl}
\affiliation{\FUW}
\author{Elena Blundo}
\affiliation{\Sapienza}
\author{Natalia Zawadzka}
\affiliation{\FUW}
\author{Salvatore Cianci}
\affiliation{\Sapienza}
\author{Diana~Vaclavkova}
\affiliation{\LNCMI}
\author{Piotr Kapu\'sci\'nski}
\affiliation{\LNCMI}
\author{Dipankar Jana}
\affiliation{\LNCMI}
\author{Giorgio Pettinari}
\affiliation{\CNR}
\author{Marco~Felici}
\affiliation{\Sapienza}
\author{Karol~Nogajewski}
\affiliation{\FUW}
\author{Miroslav Barto\v{s}}
\affiliation{\Brno}
\author{Kenji~Watanabe}
\affiliation{\Watanabe}
\author{Takashi Taniguchi}
\affiliation{\Taniguchi}
\author{Clement~Faugeras}
\affiliation{\LNCMI}
\author{Marek Potemski}
\affiliation{\FUW}
\affiliation{\LNCMI}
\affiliation{\Centera}
\author{Adam Babi\'nski}
\affiliation{\FUW}
\author{Antonio Polimeni}
\affiliation{\Sapienza}
\author{Maciej R. Molas}
\email{maciej.molas@fuw.edu.pl}
\affiliation{\FUW}

\begin{abstract}
The possibility of almost linear tuning of the band gap and of the electrical and optical properties in monolayers (MLs) of semiconducting transition metal dichalcogenide (S-TMD) alloys opens up the way to fabricate materials with on-demand characteristics. 
By making use of photoluminescence spectroscopy, we investigate optical properties of WSSe MLs with a S/Se ratio of 57/43 deposited on SiO$_2$/Si substrate and encapsulated in hexagonal BN flakes. 
Similarly to the $"parent"$ WS$_2$ and WSe$_2$ MLs, we assign the WSSe MLs to the ML family with the dark ground exciton state. 
We find that, in addition to the neutral bright A exciton line, three observed emission lines are associated with negatively charged excitons. 
The application of in-plane and out-of-plane magnetic fields allows us to assign undeniably the bright and dark (spin- and momentum-forbidden) negative trions as well as the phonon replica of the dark spin-forbidden complex. Furthermore, the existence of the single photon emitters in the WSSe ML is also demonstrated, thus prompting the opportunity to enlarge the wavelength range for potential future quantum applications of S-TMDs. 

\end{abstract}

\maketitle

\section{Introduction \label{sec:Intro}}
Semiconducting transition metal dichalcogenides (\mbox{S-TMDs}) MX$_2$, which crystallise in the 2H phase, include only five compounds, $i.e.$ WS$_2$, WSe$_2$, MoS$_2$, MoSe$_2$, and MoTe$_2$~\cite{Koperski2017}. Monolayers (MLs) of MX$_2$ are direct-band-gap semiconductors and can emit light more efficiently than their bulk counterparts having indirect band gaps.
The fundamental optical transition in S-TMD MLs, the so-called A exciton, spans the spectral range from ~1.15 eV for the MoTe$_2$ ML~\cite{Lezama2015} up to $\sim$2.1 eV for the WS$_2$ ML~\cite{Zinkiewicz2021}. 
One possibility of adjusting the A-exciton energy is to change the layer thickness, but this is accompanied with a change in the band gap character (from direct to indirect) with detrimental effects on the radiative efficiency~\cite{Arora2015, aroramose2, Lezama2015, molasNanoscale}.
Yet another strategy is that of straining S-TMDs, thus shifting the exciton energy.
However, strain also leads to direct-to-indirect exciton transitions, affecting not only the optical efficiency and carrier decay time\cite{Blundo_prr}, but also having huge effects on the exciton $g$-factor~\cite{Blundo_magneto_prl}.
Another opportunity to adjust the energy of the optical band gap is by mixing different atoms, $e.g.$ transition metals (Mo and W) or chalcogens (S, Se, and Te), which leads to the formation of ternary compounds (alloys), such as MoSSe, WSSe, MoWSe$_2$, etc. 
In such materials, it is possible to tune the optical band gap by variation of the stoichiometry ratio of each element~\cite{Sun2017, Xie2015, Zhao2018, Nugera2022}.
Using a variety of candidates, the band gap of ternary compounds can be continuously tuned within a wide spectral range.
This opens up new possibilities for absorbing and emitting light at desired wavelengths, which plays a crucial role in potential optoelectronic applications.
The S-TMD MLs are known to be organised in two subgroups, $i.e.$ $bright$ and $darkish$, due to the type of the ground exciton state (bright and dark, respectively)~\cite{Molas2017}.
In bright MLs, the optical recombination between spin-polarised subbands of the top valence band (VB) and the bottom conduction band (CB) is optically active (bright). 
In the case of darkish MLs, that transition is optically inactive (dark). 
MoSe$_2$ and MoTe$_2$ MLs are bright, while MoS$_2$, WS$_2$, and WSe$_2$ MLs are darkish~(see Refs.~\cite{Molas2019, Robert2020, Zinkiewicz2020}). 
The determination of bright or darkish character of the excitonic ground state of alloy MLs is still missing.
However, in the mean field approximation, it is expected that the electronic and optical properties of alloy MLs ($e.g.$, WSSe) are an average of those of the corresponding $"parents"$ ($e.g.$, WS$_2$ and WSe$_2$).

In this work, we investigate the optical response of MLs of WSSe with a S/Se ratio of 57/43 encapsulated in hexagonal BN (hBN) flakes, by means of photoluminescence (PL) spectroscopy.
The WSSe ML is ascribed to the family of MLs with a dark ground exciton state, like its $"parent"$ WS$_2$ and WSe$_2$ MLs. 
Our optical spectroscopy investigations furthermore reveal additional emission lines at energies lower than that of the bright excitons.
By applying magnetic fields perpendicular and parallel to the layer plane, we are able to determine that these lines arise from bright and dark (spin- and momentum-forbidden) negatively charged excitons and to identify a phonon replica of the dark spin-forbidden charged exciton complex. 
We finally demonstrate WSSe MLs as a potential platform for quantum communication as these MLs, similar to WSe$_2$ or WS$_2$~\cite{Palacios-Berraquero2016, Koperski2017, Chakraborty2019, Cianci_OpticalMaterials} host centres emitting photons one-by-one, as shown by a clear anti-bunching at zero delay in our auto-correlation measurements of their emission statistics. 

\section{Results \label{sec:results}}

\begin{figure}[t]
	\subfloat{}%
	\centering
	\includegraphics[width=1\linewidth]{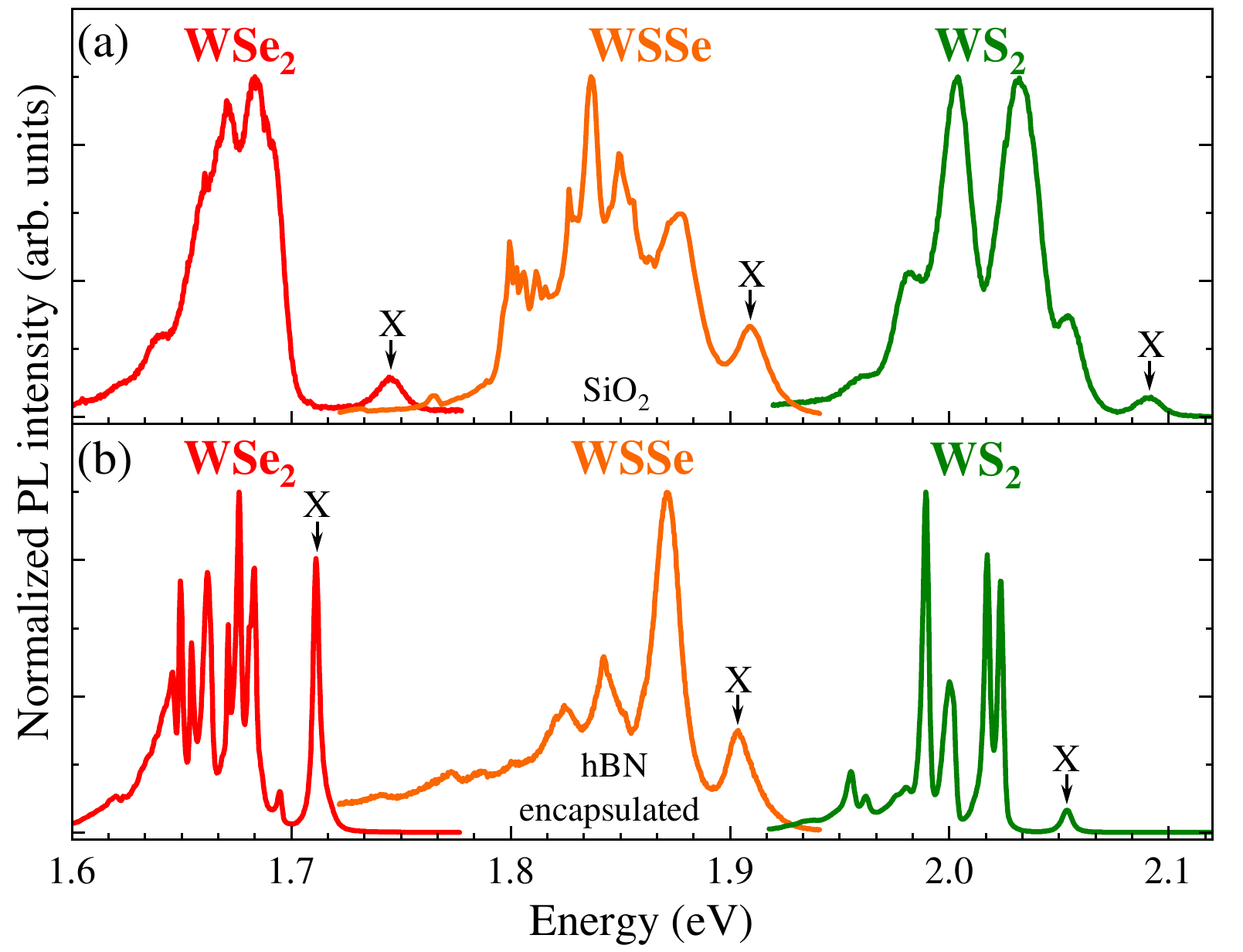}%
	\caption{Low-temperature ($T$=5~K) PL spectra of WSe$_2$, WS$_2$, and WSSe MLs (a) deposited directly onto a SiO$_2$/Si substrate and (b) encapsulated in hBN flakes. The spectra were measured under excitation energy of 2.41~eV and power of 50~$\mu$W.}
	\label{fig:pl}
\end{figure}


Low-temperature ($T$=5~K) PL spectra of WSe$_2$, WSSe, and WS$_2$ MLs deposited on SiO$_2$/Si substrates (top) and encapsulated in hBN flakes (bottom) are shown in Fig.~\ref{fig:pl}. 
The relative concentration of S and Se atoms in the WSSe crystal, from which MLs were exfoliated was determined by means of energy dispersive X-ray analysis.
The alloy was found to be homogeneous over the whole investigated crystal ($\sim$ half-mm in size) with a S/Se ratio $\approx 57/43$, see Supplementary Information (SI) for details.
First, we focus on the analysis of the PL spectra measured on MLs deposited on a SiO$_2$/Si substrate, see Fig.~\ref{fig:pl}(a).
The corresponding PL spectra of the WS$_2$ and WSe$_2$ MLs display characteristic emission lines, labelled X, associated with the recombination of neutral bright A excitons comprising carriers from the K$^\pm$ points of the Brillouin zone (BZ)~\cite{Koperski2017}.
The determined X energies for the WS$_2$ and WSe$_2$ MLs are equal to about 2.090~eV and 1.745~eV, respectively. 
By analogy, we ascribe the emission line apparent at about 1.909~eV in the WSSe ML to the A exciton. Its energy is very close to 1.942~eV, as expected for the alloy ML from the linear dependence of the X energy with the S/Se relative concentration.
As can be seen in Fig.~\ref{fig:pl}(b), the encapsulation of MLs in hBN redshifts the X energies. 
This results from the reduction of both the band gap energies and the binding energies of excitons in the hBN-encapsulated MLs, as compared to MLs deposited on a Si/SiO$_2$ substrate. 
This reduction is due to the different dielectric environments of the MLs (hBN encapsulation versus vacuum and Si/SiO$_2$ substrate)~\cite{Vaclavkova2018}.
Note that a series of low-energy emission lines can be observed in Fig.~\ref{fig:pl}. 
Their attribution for WS$_2$ and WSe$_2$ MLs was discussed in the literature~\cite{Arora2015, molasNanoscale, Liu2020, Zinkiewicz2021}, while the corresponding analysis for the WSSe ML was missing and will be presented in the following.
The most well-known effect of the hBN encapsulation is a significant reduction in the inhomogeneous broadening of the spectral lines, leading to linewidths that approach the radiative decay limit~\cite{Cadiz2017}.
To study the hBN influence in our case, we fitted the X lines, seen in Fig.~\ref{fig:pl}, with Lorentzian functions. 
Their extracted linewidths, $i.e.$ full width at half maximum (FWHM), decrease substantially from 13~meV (16~meV) for the WSe$_2$ (WS$_2$) MLs deposited on a Si/SiO$_2$ substrate to a few meV for MLs encapsulated in hBN flakes (3.5 meV for WSe$_2$ and 5.2 meV for WS$_2$).
Interestingly, the linewidth reduction for the WSSe ML (from 23~meV for Si/SiO$_2$ to 18~meV for encapsulated ML) is not as large.
The main difference may come from the intrinsic quality of the WSSe alloy crystal compared with the pure WS$_2$ and WSe$_2$ crystals.
While the S/Se composition in our WSSe crystal is relatively homogeneous at the macroscale, local variations of the chalcogen concentration are expected at the nanoscale.
The alloy disorder as the main source of intrinsic scattering strongly limits the exciton lifetime in alloy MLs, leading to a broadening of the lines. 
Consequently, it is expected that hBN-encapsulation is not able to yield to a sizeable reduction of the X linewidth. 

\begin{figure}[t]
	\subfloat{}%
	\centering
	\includegraphics[width=1\linewidth]{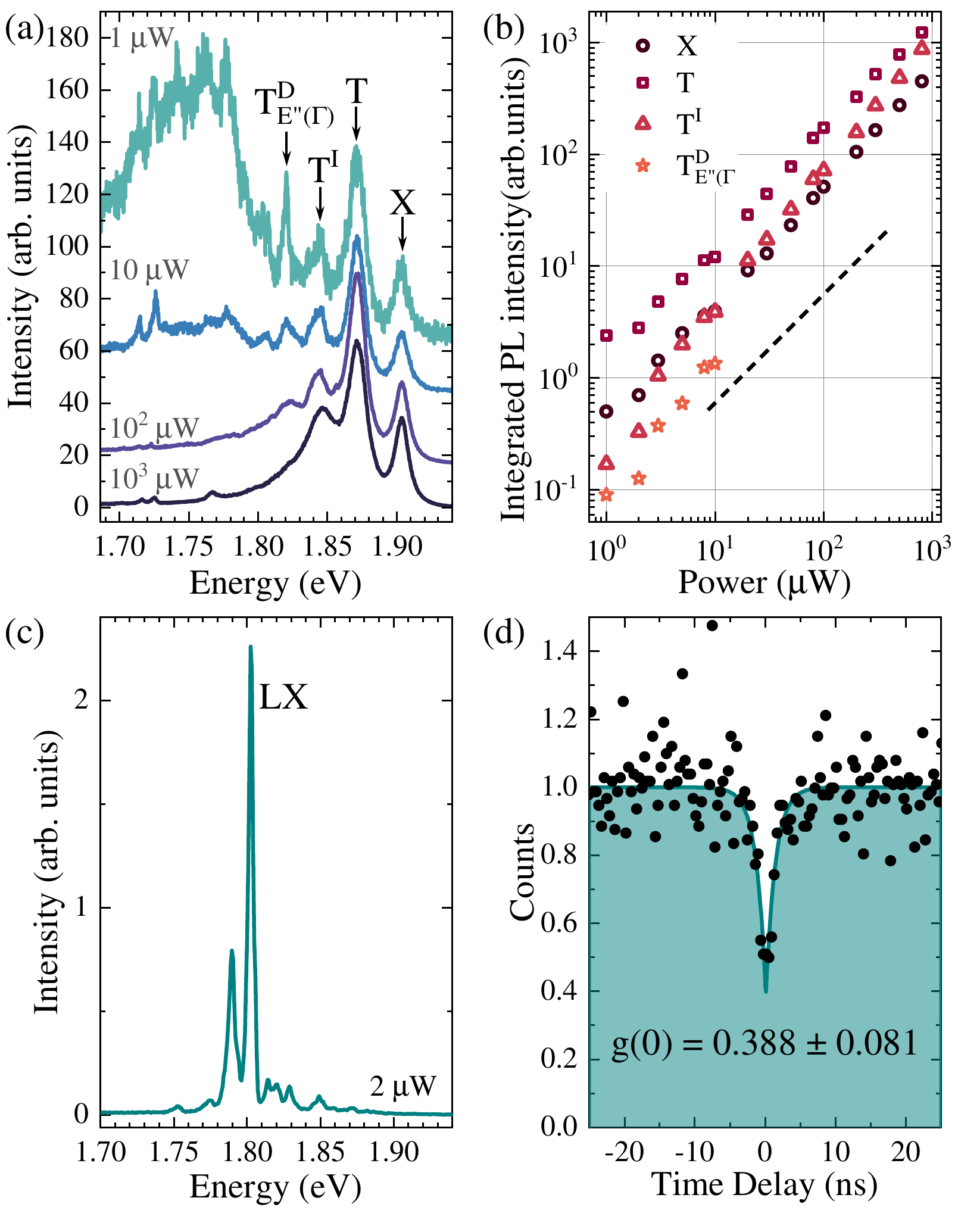}
	\caption{(a) Power dependence of low-temperature ($T$=5 K) PL spectra measured on a WSSe monolayer with 2.41 eV laser light excitation. (b) The intensity evolution of the emission features with excitation power. The dashed black line indicates the linear  dependence as a guide to the eye. (c) The low-temperature PL spectrum of a single photon emitter, denoted as LX, measured on a WSSe ML.
		Note that the presented spectrum was measured on a specific region of a WSSe ML different from the ML displayed in panel (a). (d) The photon autocorrelation histogram recorded for the LX line.}
	\label{fig:power}
\end{figure}

\begin{figure}[t]
	\subfloat{}%
	\centering
	\includegraphics[width=1\linewidth]{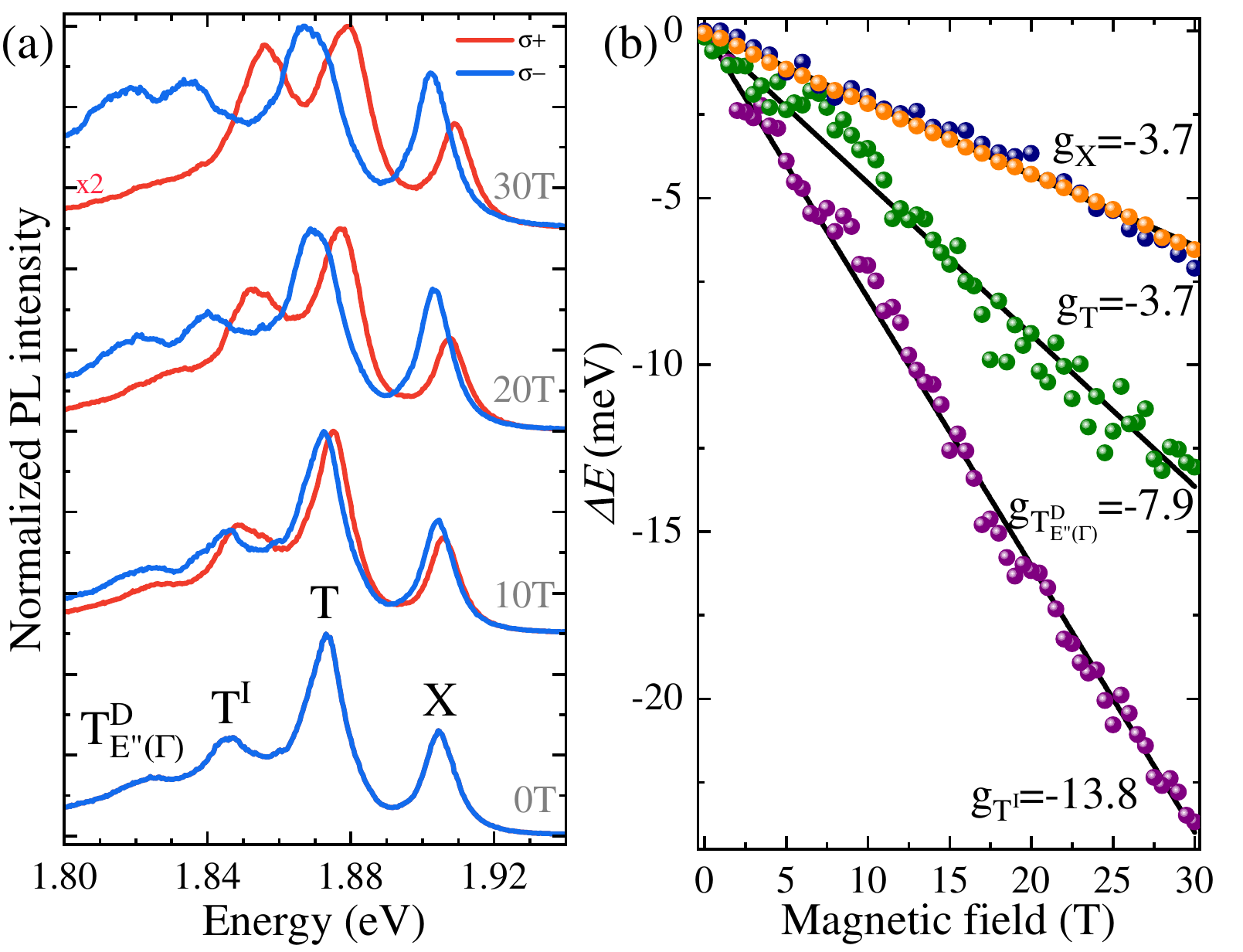}%
	\caption{(a) Helicity-resolved PL spectra of an hBN-encapsulated WSSe monolayer at $T$=4.2~K measured at selected values of the applied out-of-plane magnetic field. The red (blue) colour corresponds to the $\sigma^+$ ($\sigma^-$) polarised spectra. The spectra are vertically shifted for clarity. (b) Energy difference between the two circularly polarised split components of the X, T, T$^\textrm{I}$, and T$^{\textrm{D}}_{{\textrm{E}}"(\Gamma)}$ transitions as a function of out-of-plane magnetic field. The solid lines represent fits according to the equation described in the text. The spectra were measured under excitation energy of 2.41~eV and power of 50~$\mu$W.}%
	\label{fig:Far}
\end{figure}

The assignment of the lower-in-energy features, labelled T, T$^\textrm{I}$ and T$^\textrm{D}_{\textrm{E}"(\Gamma)}$ in Fig.~\ref{fig:power}(a), requires a more detailed analysis.
The T line can be ascribed to the negatively charged exciton (negative trion) according to the following evidence: 
(i) the energy separation between the X and T lines is of 32~meV, which is very close to the corresponding separation between the neutral exciton and the negative trion reported in the  WSe$_2$ (32~meV)~\cite{Liu2020,He2020} and WS$_2$ MLs (34~meV)~\cite{Vaclavkova2018,Zinkiewicz2020};
(ii) the T emission line can be observed up to room temperature (see SI for details), which is typical of the negative trion lines in W-based MLs~\cite{Arora2015,Molas2017,Vaclavkova2018};
(iii) the sign of free carriers can be revealed indirectly due to the apparent double structure of the T lines at the highest magnetic fields, as discussed later (see Fig.~\ref{fig:Far}(a)), which is a signature of negatively charged excitons in WSe$_2$ and WS$_2$ MLs~\cite{Liu2020, He2015, Vaclavkova2018, Zinkiewicz2021}.
The assignment of the T$^\textrm{I}$ and T$^\textrm{D}_{\textrm{E}"(\Gamma)}$ lines to the correspondingly intervalley momentum-forbidden negative dark trion and the phonon replica of the intravalley spin-forbidden dark trion, requires the use of external magnetic fields applied in different configurations with respect to the ML plane (parallel or perpendicular), which is discussed below.
To confirm the origin of the aforementioned emission lines, we investigated their intensity evolution as a function of the excitation power.
The low-temperature PL spectra of a WSSe ML encapsulated in hBN flakes measured with four different excitation powers are presented in Fig.~\ref{fig:power}(a), while the obtained power dependences of the studied emission lines are shown in Fig.~\ref{fig:power}(b). 
Note that the power evolution of the T$^\textrm{D}_{\textrm{E}"(\Gamma)}$ line could not be resolved for an excitation power greater than 300 $\mu$W. 
The integrated intensities of all investigated lines are characterised by an almost linear evolution with the excitation power.
To analyse quantitatively the power evolutions of the line intensities, we fitted the corresponding dependences using the following formula $I\propto P^\beta$ , where $I$ is integrated intensity of lines, $P$ is the excitation power, and $\beta$ is a power coefficient.
The extracted values for the slopes of these linear evolutions range from about 1 for the X, T, and  T$^\textrm{D}_{\textrm{E}"(\Gamma)}$ lines to 1.2 for the T$^\textrm{I}$, which is expected in the case of excitonic complexes composed of a single electron-hole ($e$-$h$) pair (in contrast to biexcitons, whose intensities increase quadratically with excitation power)~\cite{Klingshirn2012}.

It is important to mention that in addition to the free exciton and trion lines, a zoo of states resulting in a broad band below 1.8 eV can also be observed in Fig.~\ref{fig:power}(a). 
These emission lines feature a sublinear behaviour with power and are thus attributed to defect-related emissions. 
Interestingly, at some specific locations where the ML is wrinkled or blistered (and thus strained~\cite{Blundo_AnalyticalModel,DiGiorgio_review_elastic}) due to the trapping of contaminants or close to the edges, isolated narrow lines show up from the broad defect-related band, like those in Fig.~\ref{fig:power}(c).
These lines can be attributed to localised excitons (LXs), similar to what was reported for WSe$_2$~\cite{Koperski2017,Cianci_OpticalMaterials}.
To verify their quantum nature, we performed autocorrelation measurements on the LX line, see Fig.~\ref{fig:power}(c).
As shown in Fig.~\ref{fig:power}(d), a clear antibunching behaviour is observed, resulting in a second-order autocorrelation function $g(0) = 0.388 \pm 0.081$, thus proving the single-photon emitter nature of the LXs.
Our result confirms the promise of WSSe for quantum applications and calls for a deeper investigation of the LX properties and of the role played by strain, which will be the object of future studies.


\begin{figure}[!t]
	\subfloat{}%
	\centering
	\includegraphics[width=1\linewidth]{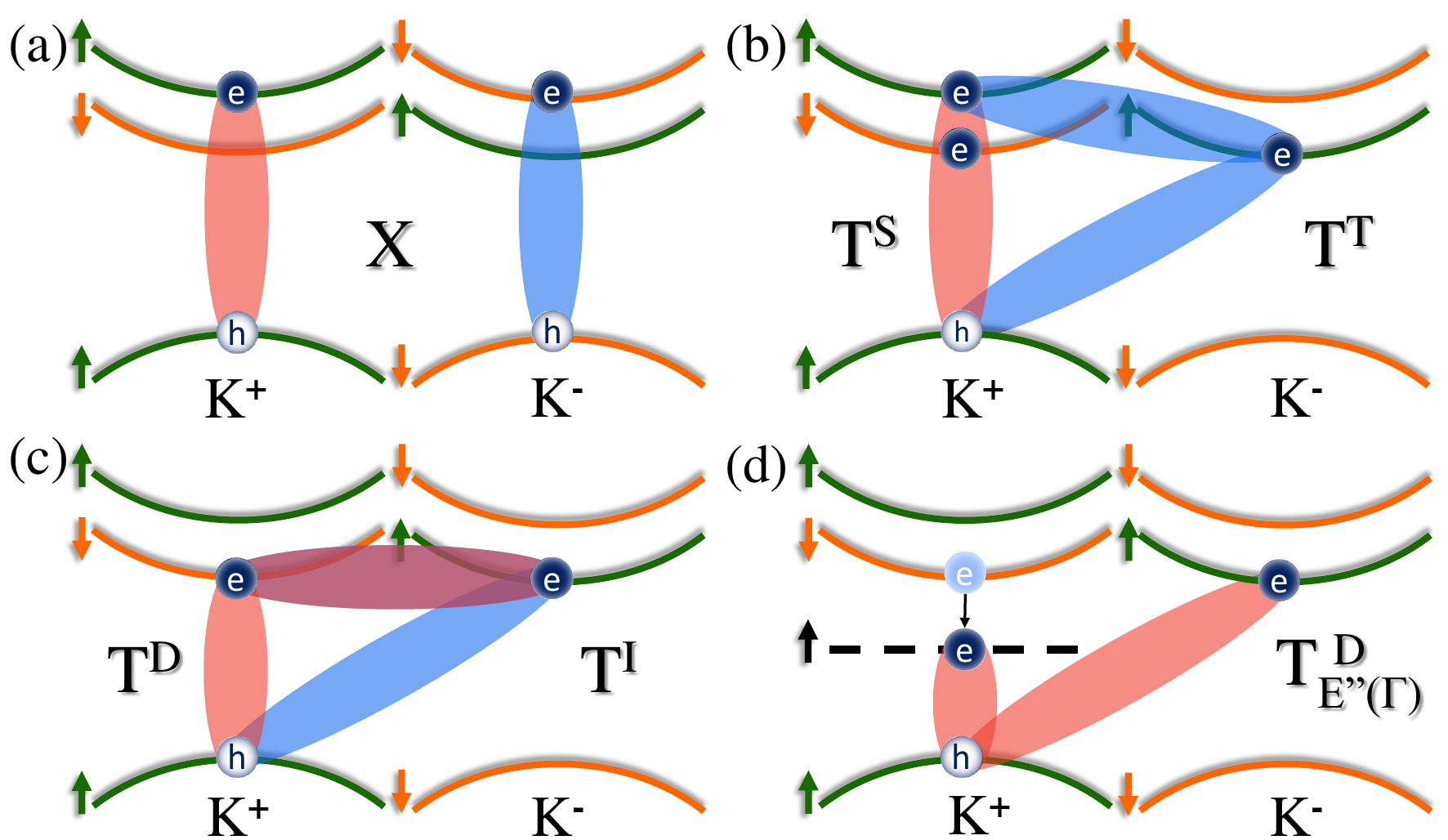}%
	\caption{Schematic illustration of possible spin configurations for (a) the neutral exciton (X), (b) the bright singlet and triplet negative trions (T$^\textrm{S}$ and T$^\textrm{T}$, respectively), (c) the dark intravalley and intervalley negative trions (T$^\textrm{D}$ and T$^\textrm{I}$, respectively), and (d) the dark trion emission assisted by the optical phonon E" from $\Gamma$ point of the BZ (T$^{\textrm{D}}_{{\textrm{E}}"(\Gamma)}$) formed in the vicinity of optical band gap of WSSe ML. Note that we draw only complexes for which a hole is located at the K$^+$ point of the BZ.}
	\label{fig:bands}
\end{figure}

\begin{figure}[t]
	\subfloat{}%
	\centering
	\includegraphics[width=1\linewidth]{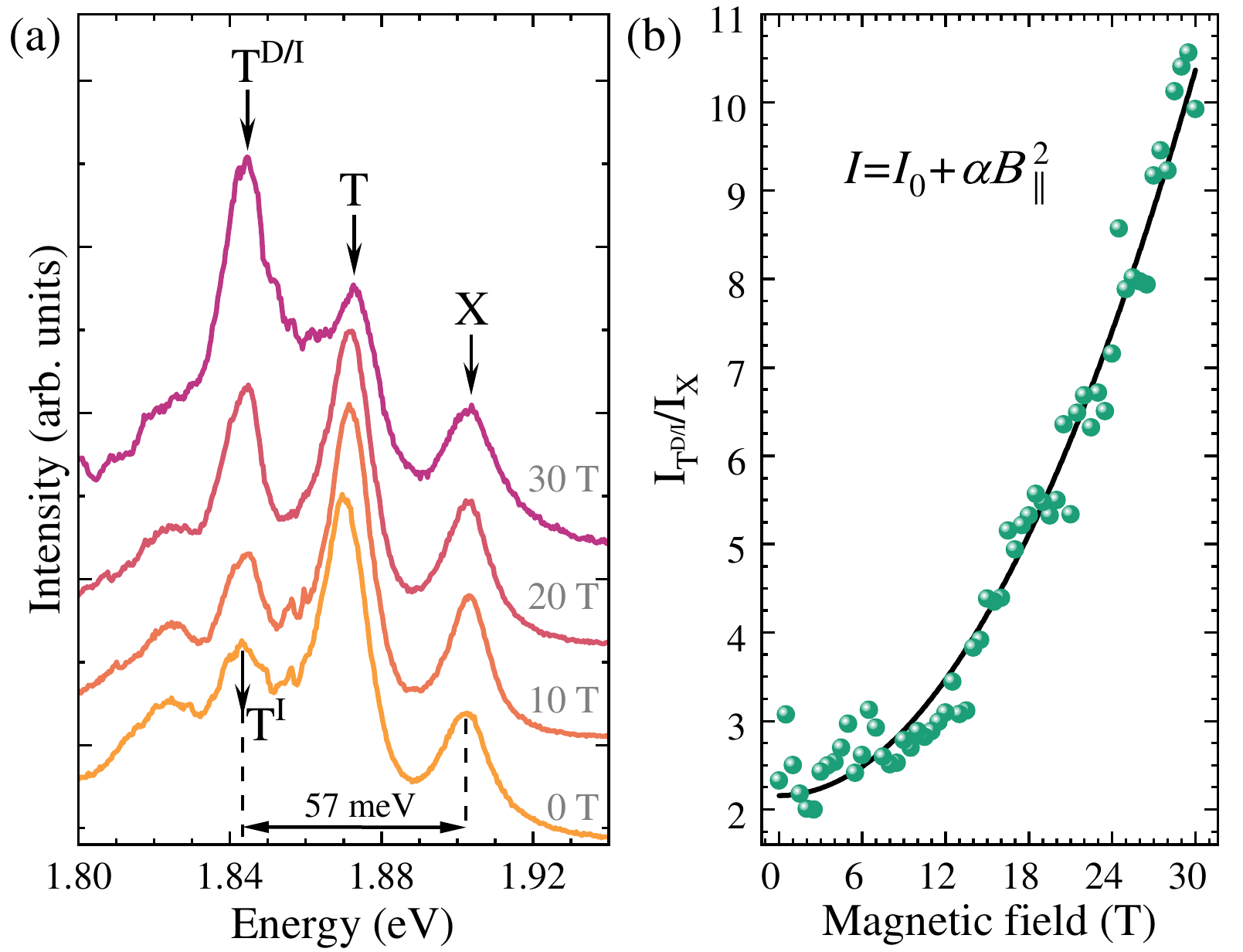}%
	\caption{(a) Low-temperature ($T$=4.2~K) PL spectra of a WSSe monolayer encapsulated in hBN flakes measured at selected magnetic fields applied in the ML's plane. The PL spectra are normalised to the intensity of the X line. (b) The corresponding magnetic-ﬁeld dependence of the relative intensity of the dark trion line to the neutral exciton line, I$_{\textrm{T}^\textrm{I/D}}$/I$_\textrm{X}$. The solid black curve represents a quadratic ﬁt according to the equation described in the text. The spectra were measured under excitation energy of 2.41~eV and power of 50~$\mu$W.}
	\label{fig:Voight}
\end{figure}

From the analysis presented so far, we can conclude that the low-temperature PL spectra of the WSSe ML encapsulated in hBN flakes comprise emission lines due to excitonic complexes composed of an $e$-$h$ pair.
Fig.~\ref{fig:bands} shows possible spin configurations for the bright and dark excitons formed in the vicinity of the A exciton.
The neutral bright exciton in the K$^\pm$ valley is composed of an electron from the higher-lying level of the conduction band (CB) and a hole from the top level of the valence band (VB) from the same K$^\pm$ point, see Fig.~\ref{fig:bands}(a).
The negatively charged exciton (negative trion) is a three-particle complex composed of an $e$-$h$ pair and an excess electron.
There are four negative trions in darkish MLs based on tungsten (W) atoms (WS$_2$ or WSe$_2$)~\cite{Liu2020, He2020, Zinkiewicz2021}, $i.e.$ two bright and two dark states.
These states can be formed at the K$^+$ and K$^-$ points due to the location of a hole, leading to two possible configurations of a given complex. 
Note that only a single configuration of a specific complex is shown in panels (b), (c), and (d) of Fig.~\ref{fig:bands}.
Due to the spin conservation rule for S-TMD MLs, the bright (optically active) negative trion can be found in both the intravalley singlet (T$^\textrm{S}$), involving two electrons from the same valley, and the intervalley triplet (T$^\textrm{T}$), comprising two electrons from different valleys~\cite{Vaclavkova2018}.
For dark (optically inactive) negative trions, the corresponding electrons are located in different valleys and characterised by antiparallel alignment of their spins.
This configuration leads to two complexes, depending on the state of electron associated with the recombination process: intravalley spin-forbidden (T$^\textrm{D}$) and intervalley momentum-forbidden (T$^\textrm{I}$), which cannot recombine optically due to spin and momentum conservation, respectively~\cite{LiuGate2019, Zinkiewicz2020, Zinkiewicz2021}.
One of the possibilities to fulfil the spin and momentum conservation rules during optical recombination of dark trions is phonon emission from the $\Gamma$ and K points of the BZ, respectively~\cite{LiuValley, Liu2020, He2020, Zinkiewicz2021}.
Fig. ~\ref{fig:bands}(d) presents the schematic illustration of a possible recombination pathway of dark negative trions involving optical phonon emission from the $\Gamma$ point of the BZ, T$^{\textrm{S}}_{{\textrm{E}}"(\Gamma)}$.
To verify the assignment of the T, T$^\textrm{I}$, and T$^\textrm{D}_{\textrm{E}"(\Gamma)}$ lines, we measured PL spectra as functions of in-plane and out-of-plane magnetic fields up to 30~T, see Figs.~\ref{fig:Voight} and \ref{fig:Far}.



To investigate the effect of the in-plane magnetic field ($B_\parallel$) on the emission of the WSSe ML, we measured the evolution of the low-temperature ($T$=4.2~K) PL spectra under fields up to $B_\parallel$=30~T.
Fig.~\ref{fig:Voight}(a) presents the PL spectra at selected magnetic fields, $B_\parallel$=0, 10, 20, and 30 T.
With increasing $B_\parallel$, the intensity of the T$^\textrm{I}$ line increases significantly, while the intensities of the other emission lines, seen in the PL spectra, remain almost unchanged.
The energy separation between the X and T$^\textrm{I}$ lines of approximately 57 meV is very similar to the reported energy separations between the bright neutral exciton and the spin-forbidden negative trion in both WSe$_2$ (54-60~meV)~\cite{Liu2020, Liu2020, He2020, Yang2022} and WS$_2$ (57~meV)~\cite{Zinkiewicz2021}.
Consequently, the brightening of the emission line in the in-plane field, labelled T$^\textrm{D}$, is ascribed to the intravalley spin-forbidden dark negative trion.
To support further our attribution, we analyse the magnetic field evolution of the T$^\textrm{D/I}$ intensity.
The $B_\parallel$ evolution of the intensity of the dark trion $T^D$ is expected to be quadratic $I=\alpha B_\parallel^2$~\cite{Molas2017, Zinkiewicz2021, Zinkiewicz2022}.
Fig.~\ref{fig:Voight}(b) displays the $B_\parallel$ dependence of the relative intensities of the T$^\textrm{I/D}$ and X lines in magnetic fields up to 30 T.
Note that the division by the X intensity allows us to eliminate the variation of the signal intensity during measurements, $e.g.$ the measured signal in the magnetic field setup is affected by the Faraday effect.
The extracted data are accompanied by quadratic fits with the formula $I=I_0+\alpha B_\parallel^2$, where $I_0$ corresponds to the T$^\textrm{I}$ intensity at zero field.
As can be appreciated in Fig.~\ref{fig:Voight}(b), the obtained data can be described well by the proposed formula with the $\alpha$ parameter equal to $9.1\times10^{-3}/\textrm{T}^{2}$.


The last part of this work is devoted to the effects of out-of-plane magnetic fields ($B_\perp$) on the properties of the studied lines.
Fig.~\ref{fig:Far}(a) demonstrates the helicity-resolved PL spectra measured on the hBN-encapsulated WSSe ML in magnetic fields up to 30 T oriented perpendicularly to the ML plane.
Due to both the large binding energies of free excitons and their huge reduced masses in S-TMD MLs~\cite{Stier2016, Goryca2019, Molas2019Energy},  applying the out-of-plane magnetic field results mostly in the exciton Zeeman effect~\cite{Koperski2019} (the diamagnetic shift of the ground excitonic states can hardly be seen in magnetic fields as high as 70~T~\cite{Stier2016, Goryca2019}), which manifests itself as a splitting of the two counter-circularly-polarised components of a
given transition ($\sigma^\pm$).
As can be seen in Fig.~\ref{fig:Far}(a), all the observed emission lines split into two $\sigma^\pm$ components, but the magnitude of the Zeeman splittings differs, providing us with a tool to distinguish between the different complexes.
Note that a double structure of the $\sigma^+$ component of the T line can be appreciated at 30~T with the determined energy separation between the T$^\textrm{S}$ and T$^\textrm{T}$ components of about 6 meV (see SI for details), which is in very good agreement with the previously reported corresponding energy splitting of the spin-singlet and spin-triplet negative trions in the WS$_{0.6}$Se$_{1.4}$~\cite{Meng2019}, WSe$_2$ (6-7 meV)~\cite{Courtade2017, Chen2018, Li2018}, and WS$_2$ (7-8 meV)~\cite{Vaclavkova2018, Jakubczyk2018} MLs.
The energy separation of the $\sigma^\pm$ components in the $B_\perp$ field, $\Delta E (B_\perp) = E_{\sigma^+} - E_{\sigma^-}$ can be expressed as $\Delta E (B_\perp) = g \mu_B B_\perp$, where $g$ denotes the effective $g$-factor of the considered excitonic complex and $\mu_B$ is the Bohr magneton.
The magnetic field evolutions of $\Delta E$ with linear fits to the experimental data for the X, T, T$^\textrm{I}$, and T$^\textrm{D}_{\textrm{E}"(\Gamma)}$ lines are shown in Fig.~\ref{fig:Far}(b).
The extracted $g$-factors for both the X and T lines are about -3.7.
Simultaneously, the $g$-factors found for the T$^\textrm{I}$ and T$^\textrm{D}_{\textrm{E}"(\Gamma)}$ lines are much larger and correspond to approximately \mbox{-13.8} and -7.9.
The T$^\textrm{D}_{\textrm{E}"(\Gamma)}$ assignment to the phonon replica of the T$^\textrm{D}$ line is due to its $g$-factor of about -8, which is a characteristic value of spin-forbidden transitions (see Refs.~\cite{LiuValley, He2020, Zinkiewicz2021} for details). As this type of transition requires spin-flip process of an electron in the CB (see Fig.~~\ref{fig:bands}(d)), it can be only achieved by a chiral phonon of a specific symmetry, that is E” from the $\Gamma$ point of the BZ.~\cite{LiuValley, He2020, Zinkiewicz2021} Moreover, the attribution of the T$^\textrm{D}_{\textrm{E}"(\Gamma)}$ to the emission of the E"($\Gamma$) phonon allows us to determine its energy to to be about 28 meV.
Note that the E"($\Gamma$) phonon can not be observed in a typical experimental back-scattering geometry~\cite{Duan2016, Alireza2022}, which requires that the electric field of the incident wave is perpendicular to the ML plane.
Moreover, due to the identification of the phonon replica involving the E” phonon from the $\Gamma$ point of the BZ, its energy was determined to be about 28 meV.
If we consider the corresponding E”($\Gamma$) energies in WS$_2$ (36 meV)~\cite{Zinkiewicz2021} and WSe$_2$ (21 meV)~\cite{Yang2022} MLs, the determined value is very close to the estimate of 30 meV obtained by making the assumption of a linear dependence of the phonon energy on the S/Se relative concentration
in the ML alloys.
Consequently, as reported in the literature, excitonic complexes in S-TMD MLs can be arranged into three groups due to their Zeeman splitting magnitude~\cite{Zinkiewicz2021, Liu2020, He2020, Koperski2019}:
(i) $g$-factors around -4 for bright transitions (X and T: -3.7); 
(ii) spin-forbidden dark transitions are described by the $g$-factor equal to around -8 (T$^\textrm{D}_{\textrm{E}"(\Gamma)}$: -7.9); 
and (iii) values of $g$-factors of about -14 are characteristic for momentum-forbidden dark transitions (T$^\textrm{I}$: -13.8).

\section{Summary \label{sec:summary}}
To conclude, we presented based on PL experiments investigation of the optical response of MLs of WSSe with a S/Se ratio of 57/43 deposited on Si/SiO$_2$ substare and encapsulated in hBN flakes.
We found that the WSSe ML was characterised by a dark ground exciton state, like its $"parent"$ WS$_2$ and WSe$_2$ MLs.
Moreover, the existence of single photon emitters in the WSSe ML was also demonstrated. 
Through the application of in-plane and out-of-plane magnetic fields, the four main emission lines apparent in the low-temperature PL spectra were ascribed to the neutral bright exciton, the bright and dark (spin- and momentum-forbidden) negatively charged excitons, and the phonon replica of the dark spin-forbidden trion.
Our results show that alloyed S-7TMD MLs can represent a powerful material platform to widen the spectral range of operation of 2D crystals for optoelectronics and quantum technology applications.

\section{Methods \label{sec:method}}
MLs of WSSe were mechanically exfoliated from the bulk crystals grown by the Flux zone method and purchased from 2D semiconductors.The MLs were exfoliated with a scotch tape onto polydimethylsiloxane (PDMS). 
Relatively thick hBN flakes were exfoliated mechanically onto SiO$_2$/Si substrates (300-nm-thick SiO$_2$ on top of a 500-$\mu$m-thick Si). 
The WSSe MLs were then deterministically transferred from the PDMS to the hBN flake.
Thin hBN flakes were exfoliated onto PDMS and deposited on the WSSe ML to encapsulate it. After each deposition step, the sample was annealed in high vacuum (10$^{-6}$ mbar) at $\sim120^\circ$C for several hours to induce the coalescence of the contaminants, thus improving the adhesion of the sample.
For the WSSe MLs on SiO$_2$/Si substrates, the MLs were first isolated onto PDMS and then deposited on the substrate.
The investigated MLs of WSe$_2$ and WS$_2$ ML encapsulated in hBN flakes were fabricated in an analogue way by mechanical exfoliation based on PDMS, but the whole structure was deposited on a bare Si substrate.
The WSe$_2$ and WS$_2$ MLs on SiO$_2$/Si substrates were fabricated by simple exfoliation of the bulk crystals onto the substrates.

The WSSe alloy was examined by energy dispersive X-ray analysis (EDX) using a ZEISS-Sigma300 scanning electron microscope (SEM) equipped with an Oxford Instruments X-Act 100 mm energy-dispersive spectrometer.
Data were acquired with an acceleration voltage of 28 kV and analysed by INCA software. The spatial resolution was in the 10-15 $\mu$m range.

The PL experiments at zero magnetic field were performed using a $\lambda= $514.5 nm (2.41 eV) continuous wave (CW) laser diode.
The studied samples were placed on a cold finger in a continuous flow cryostat mounted on $x$-$y$ motorised positioners.
The excitation light was focused by means of a 100x long-working-distance objective with a 0.55 numerical aperture producing a spot of about 1 $\mu$m diameter.
The signal was collected via the same microscope objective, sent through a 0.75 m monochromator, and then detected using a liquid nitrogen cooled charge-coupled device (CCD) camera. 

Low-temperature micro-magneto-PL experiments were performed in the Voigt and Faraday geometries, $i.e.$ magnetic field orientated parallel and perpendicular with respect to ML’s plane, respectively.
Measurements (spatial resolution $\sim$1 $\mu$m) were carried out with the aid of a resistive magnetic coil producing fields up to 30 T using a free-beam-optics arrangement. 
The sample was placed on top of a $x$-$y$-$z$ piezo-stage kept at $T$=4.2~K and was excited using a CW laser diode with 515 nm wavelength (2.41 eV photon energy).
The emitted light was dispersed with a monochromator of a 0.5 m focal length and detected with a CCD camera. 
The combination of a quarter-wave plate and a linear polariser was used to analyse the circular polarisation of signals (the measurements were performed with a fixed circular polarisation, whereas reversing the direction of magnetic field yielded the information corresponding to the other polarisation component due to time-reversal symmetry). 

Autocorrelation measurements were performed in a Hanbury-Brown-Twiss (HBT) setup by using a frequency-double Nd:YAG continuous wave laser emitting at $\lambda$=532.2 nm.

5\section*{Acknowledgements \label{Acknowledgements}}
The work has been supported by the National Science Centre, Poland (grants no. 2017/27/B/ST3/00205 and 2018/31/B/ST3/02111), EU Graphene Flagship Project, and the CNRS via IRP "2DM" project. 
We acknowledge the support of the LNCMI-CNRS, member of the European Magnetic Field Laboratory (EMFL). 
The Polish participation in EMFL is supported by the DIR/WK/2018/07 Grant from Polish Ministry of Education and Science.
We acknowledge support by the European Union's Horizon 2020 research and innovation programme through the ISABEL project (no. 871106).
E.B. acknowledges support from La Sapienza through the grant Avvio alla Ricerca 2021 (grant no. AR12117A8A090764).
This project was funded within the QuantERA II Programme that has received funding from the European Union’s Horizon 2020 research and innovation programme under Grant Agreement No 101017733, and with funding organisations Ministero dell'Universit\'{a} e della Ricerca (MUR) and Consiglio Nazionale delle Ricerche (CNR).
M.B. acknowledge support by the ESF under the project CZ.02.2.69/0.0/0.0/20\_079/0017436.
C.F. acknowledges support from Graskop project ANR-19-CE09-0026.
K.W. and T.T. acknowledge support from JSPS KAKENHI (grants no. 19H05790, 20H00354 and 21H05233).

\bibliographystyle{apsrev4-2}
\bibliography{biblio}

\newpage
\onecolumngrid
\setcounter{figure}{0}
\setcounter{section}{0}
\renewcommand{\thefigure}{S\arabic{figure}}
\renewcommand{\thesection}{S\Roman{section}}
	\begin{center}
	{\large{{\bf  \textsc{Electronic Supporting Information}} \\ Excitons and trions in WSSe monolayers}}
	\end{center}


\section{Analysis of the S and S\lowercase{e} atomic composition of WSS\lowercase{e} by energy dispersive X-ray analysis}

\begin{figure}[h!b]
	\subfloat{}%
		\centering
		\includegraphics[width=0.7\linewidth]{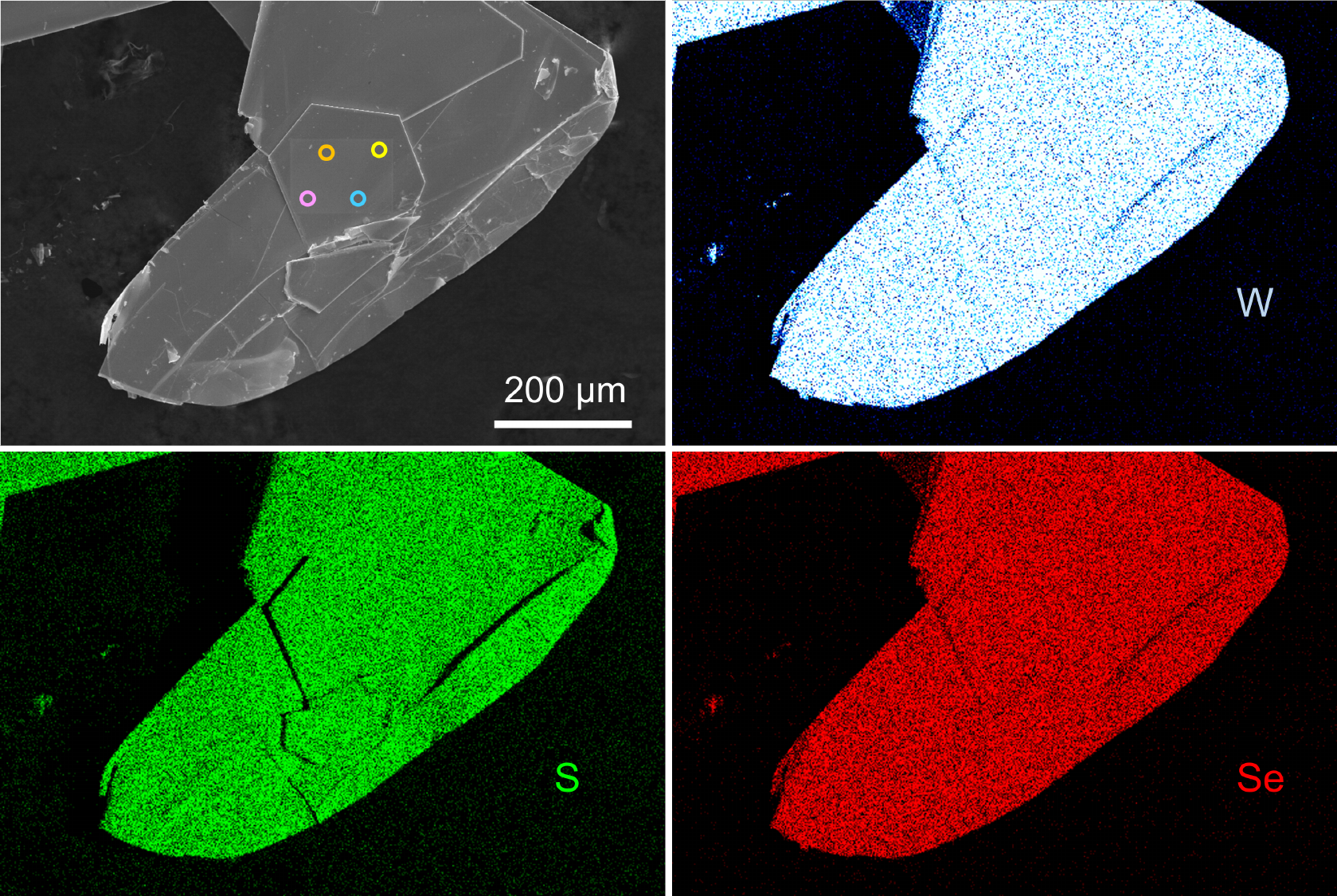}%
     	\caption{SEM-EDX map of a half-mm big piece of WSSe. The top left panel shows the electron image of the crystal, while the other panels show the elemental mapping analysis of W, S and Se (each map is normalized to its maximum intensity).}
		\label{fig:EDX_map}
\end{figure}

The WS$_\mathrm{x}$Se$_\mathrm{1-x}$ crystal used in this work was characterised by a nominal composition $\mathrm{x} = 0.5$. To verify the real composition of the sample and its uniformity over large (hundreds of $\mu$m) scales, we performed scanning electron microscopy (SEM) with energy dispersive X-ray analysis (EDX) (see Methods for details). Figure \ref{fig:EDX_map} shows the SEM image (top-left panel) of a half-mm size crystal of WSSe along with the elemental mapping analysis of W, S and Se. Indeed, a quite uniform intensity is found over the whole crystal, demonstrating that the alloys grow uniformly across large scales and that no agglomerates of variable composition are found.

\begin{figure}[!t]
	\subfloat{}%
		\centering
		\includegraphics[width=0.9\linewidth]{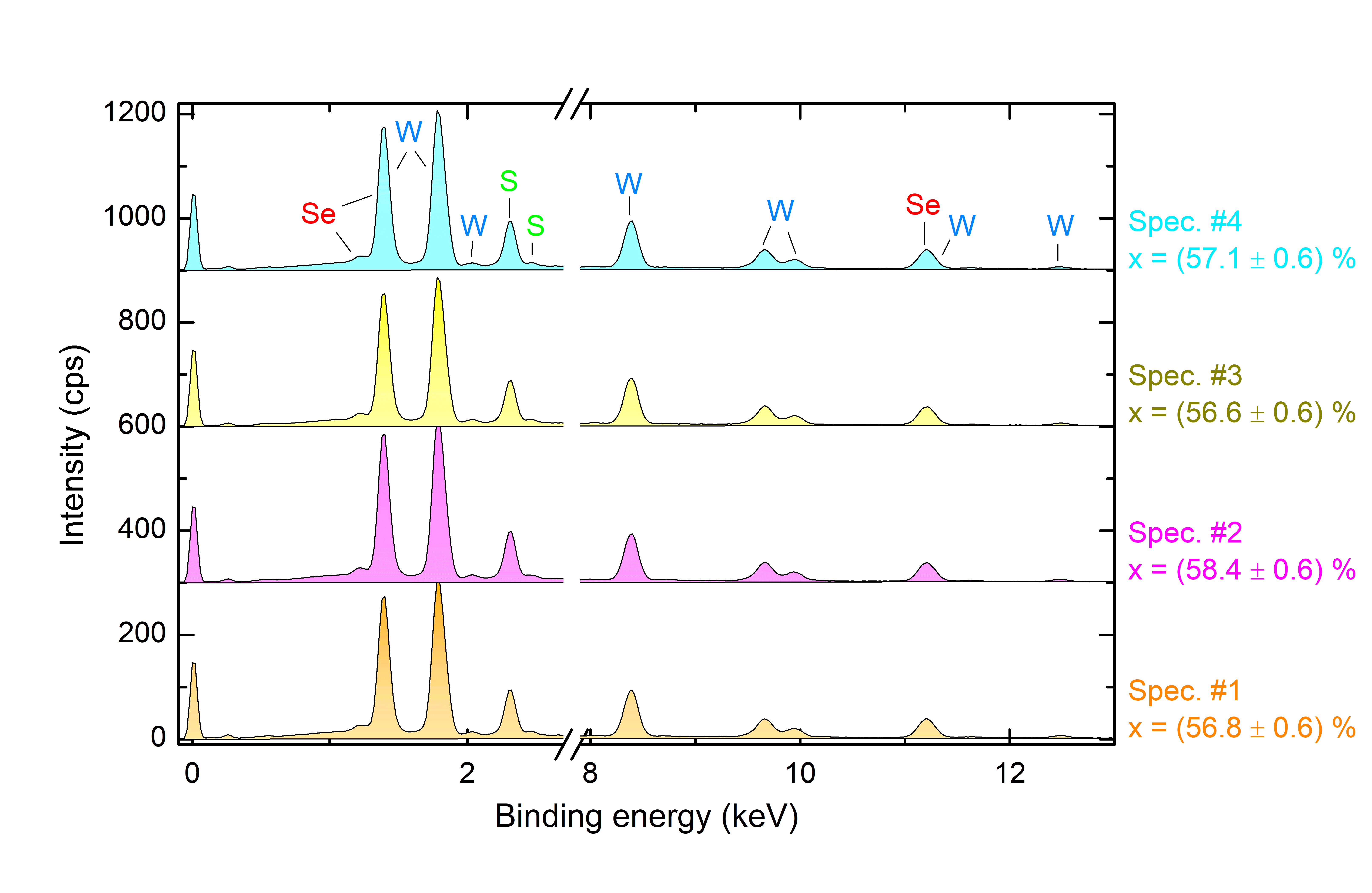}%
     	\caption{EDX high resolution spectra acquired on 4 different positions (highlighted by coloured circles in the SEM image of Figure \ref{fig:EDX_map}) over the WSSe crystal. The spectra are stacked by a constant y-offset (of 300 counts) for ease of comparison. From a quantitative analysis of the S and Se lines, we estimated the relative S composition (x) displayed on the right.}
		\label{fig:EDX_spectra}
\end{figure}

To have precise information on the sample composition, we took highly-resolved EDX spectra on 4 different points of the crystal, as shown in Fig.~\ref{fig:EDX_spectra}. From a quantitative elemental analysis, we derived the x values displayed on the right. Indeed, a relatively small variability is observed, with x being about 57 $\%$.

\clearpage

\section{Temperature evolution of the PL spectra measured on the WSS\lowercase{e} ML}

\begin{figure}[!h]
	\subfloat{}
		\centering
		\includegraphics[width=0.7\linewidth]{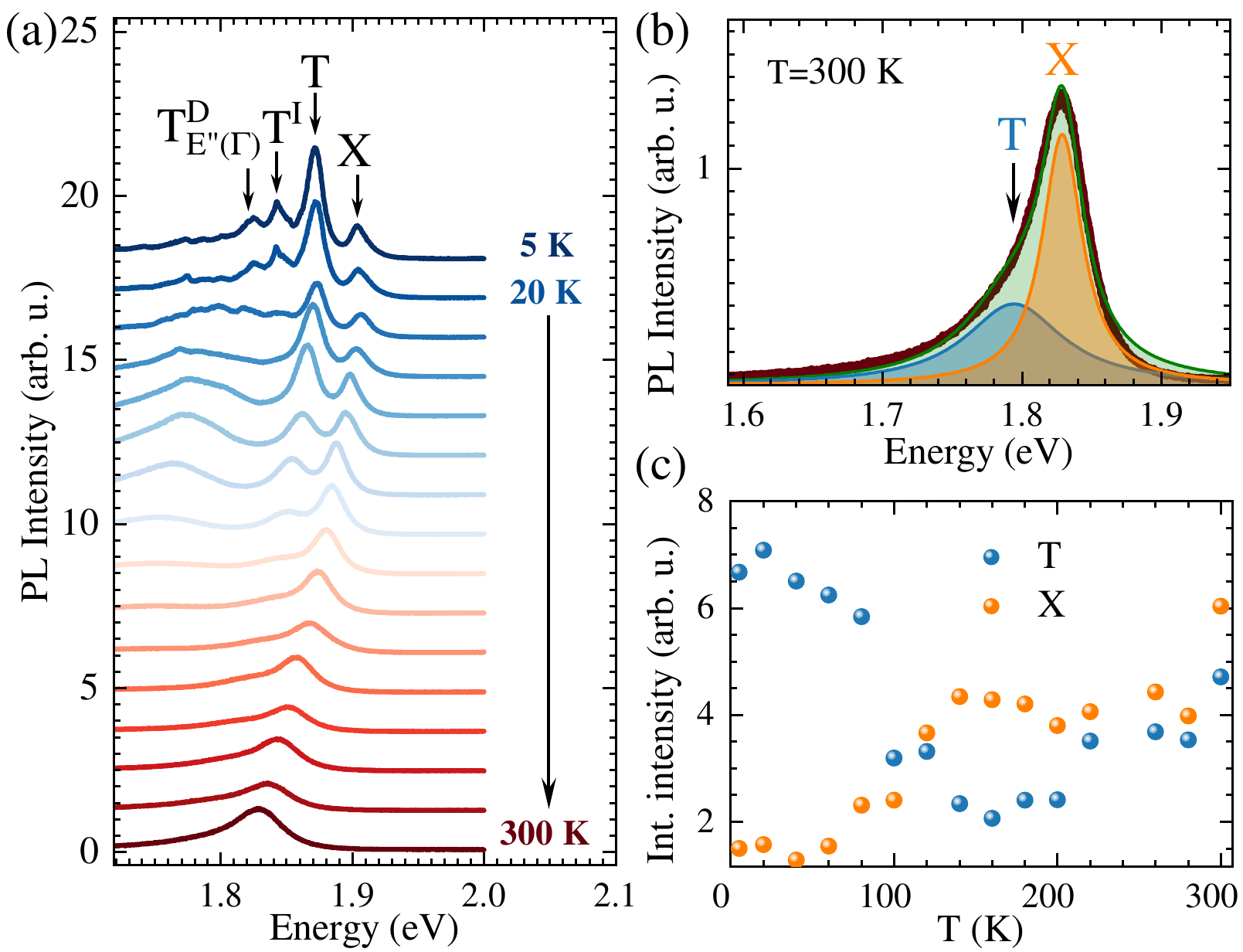}%
    	\caption{(a) Temperature evolution of PL spectra measured on ML WSSe encapsulated in hBN with 2.41 eV laser light excitation (b) The room temperature PL spectrum of the investigated WSSe ML. The coloured Lorentzian curves (orange and blue) display fits to the corresponding X and T lines. A green curve corresponds to cumulative fit. (c) Temperature evolutions of the integrated intensities of X and T emission lines.}
		\label{fig:temp}
\end{figure}

The PL spectra measured at moderate excitation power (50 $\mu$W) as a function of temperature are shown in Fig.~\ref{fig:temp} (a). As it was reported in Refs.~\citenum{Molas2017Nanoscale, Vaclavkova2018, Jadczak2017}, the low-energy peaks denoted as T$^\textrm{I}$ and T$^\textrm{D}_{\textrm{E}"(\Gamma)}$ progressively disappear one after another from the spectrum with the increased temperature. 
The neutral exciton emission dominates the spectra in the limit of high (room) temperature. The trion emission can be also distinguished up to room temperature.  The room temperature PL spectrum of the WSSe ML is shown in Fig.~\ref{fig:temp} (b). The spectrum can be well deconvoluted using two Lorentzian curves. Fig.~\ref{fig:temp7} (c) presents the temperature evolution of the integrated intensity of the X and T emission lines. It is seen that the T intensity decreases a few times with temperature increased from 10~K to 300~K, while the corresponding X intensity grows significantly (of about 6 times). The obtained temperature dependence of the neutral exciton is characteristic for the monolayers of the semiconducting transition metal dichalcogenides with the dark ground excitonic state ($e.g.$ WSe$_2$), see Refs.~\citenum{Zhang2015,Wang2015}.

\newpage

\section{The $\sigma^+$-polarised PL spectrum at out-of-plane field of 30~T}

\begin{figure}[!h]
		\subfloat{}%
		\centering
		\includegraphics[width=0.65\linewidth]{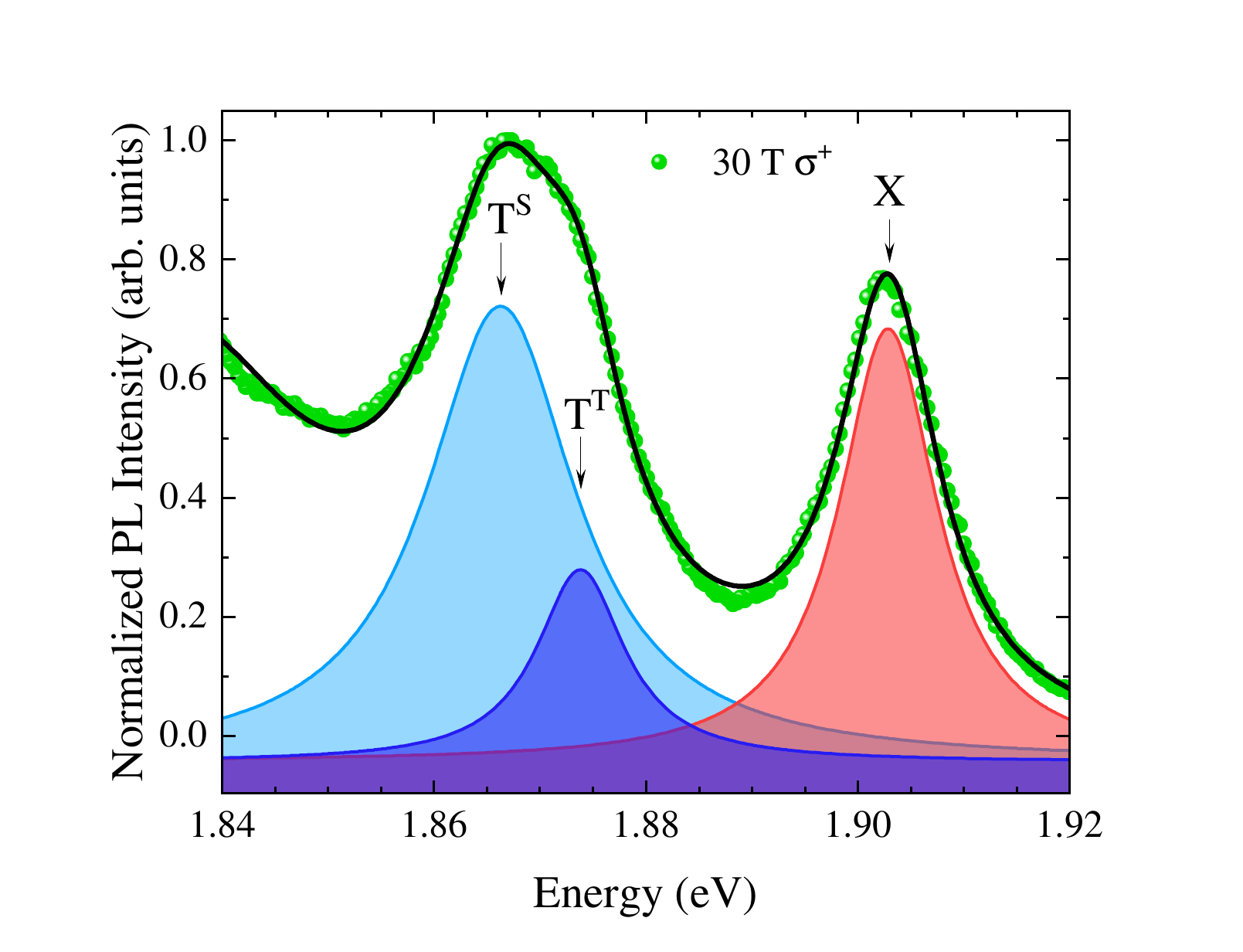}%
    	\caption{$\sigma^+$-polarised PL spectrum of WSSe monolayer measured at out-of-plane magnetic filed of 30 T. The coloured Lorentzian curves display fits to the X, T$^\textrm{T}$ and T$^\textrm{S}$ lines. A black curve corresponds to cumulative fit.}%
		\label{fig:trion1}
\end{figure}

In order to analyse the fine structure of the trion line in detail, we fitted the $\sigma^+$-polarised PL spectrum with a set of Lorentz functions measured in the out-of-plane field of 30~T, see Fig.~\ref{fig:trion1}. 
The double structure of the trion line, denoted as T$^\textrm{T}$ and T$^\textrm{S}$, can be clearly seen in the Figure. 
The extracted energy separation between the T$^\textrm{S}$ and T$^\textrm{T}$ is about 6 meV, which is in very good agreement with the previously reported corresponding energy splitting of the spin-singlet and spin-triplet negative trions in the WSe$_2$ (6-7 meV)~\cite{Courtade2017, Chen2018, Li2018} and WS$_2$ (7-8 meV)~\cite{Vaclavkova2018, Jakubczyk2018} MLs. 

\end{document}